%% file: paper.tex
\documentclass[sigplan,screen,nonacm]{acmart}

\usepackage{amsmath,amsfonts}
\usepackage{algorithmic}
\usepackage{graphicx}
\usepackage{textcomp}
\usepackage{xcolor}
\usepackage{listings, listings-rust}
\usepackage{xspace}
\usepackage{url}

\usepackage{pgfplots}
\pgfplotsset{compat=1.18}
\usepackage{pgfplotstable}
\usepackage{booktabs}

\newcommand{\cudaoxide}{\texttt{cuda-oxide}\xspace}

\usepackage{tikz}
\usetikzlibrary{arrows.meta, positioning, fit, calc, patterns.meta, shapes.multipart}
\usepackage{adjustbox}

\def\BibTeX{{\rm B\kern-.05em{\sc i\kern-.025em b}\kern-.08em
    T\kern-.1667em\lower.7ex\hbox{E}\kern-.125emX}}

\begin{document}

\title{GPU Offload in Rust: Portable, Safe, and Fast}

\author{Manuel S. Drehwald}
\email{manuel.drehwald@utoronto.ca}
\affiliation{%
  \institution{University of Toronto}
  \city{Toronto}
  \country{Canada}
}
\affiliation{%
  \institution{Lawrence Livermore National Laboratory}
  \city{Livermore}
  \state{California}
  \country{USA}
}
\additionalaffiliation{%
  \institution{Vector Institute for Artificial Intelligence}
  \city{Toronto}
  \country{Canada}
}

\author{Marcelo Domínguez}
\email{ma.dominguez.2022@alumnos.urjc.es}
\affiliation{%
  \institution{Universidad Rey Juan Carlos}
  \city{Madrid}
  \country{Spain}
}
\author{Kevin Sala}
\email{salapenades1@llnl.gov}
\affiliation{%
  \institution{Lawrence Livermore National Laboratory}
  \city{Livermore}
  \state{California}
  \country{USA}
}
\author{Alán Aspuru-Guzik}
\email{alan@aspuru.com}

\affiliation{%
  \department{Department of Computer Science}
  \institution{University of Toronto}
  \city{Toronto}
  \state{Ontario}
  \country{Canada}
}

\additionalaffiliation{%
  \department{Department of Chemistry}
  \institution{University of Toronto}
  \city{Toronto}
  \state{Ontario}
  \country{Canada}
}

\additionalaffiliation{%
  \department{Department of Materials Science \& Engineering}
  \institution{University of Toronto}
  \city{Toronto}
  \state{Ontario}
  \country{Canada}
}

\additionalaffiliation{%
  \department{Department of Chemical Engineering \& Applied Chemistry}
  \institution{University of Toronto}
  \city{Toronto}
  \state{Ontario}
  \country{Canada}
}

\additionalaffiliation{%
  \department{Institute of Medical Science}
  \institution{University of Toronto}
  \city{Toronto}
  \state{Ontario}
  \country{Canada}
}

\additionalaffiliation{%
  \institution{Vector Institute for Artificial Intelligence}
  \city{Toronto}
  \state{Ontario}
  \country{Canada}
}

\additionalaffiliation{%
  \institution{Acceleration Consortium}
  \city{Toronto}
  \state{Ontario}
  \country{Canada}
}

\additionalaffiliation{%
  \institution{Canadian Institute for Advanced Research (CIFAR)}
  \city{Toronto}
  \state{Ontario}
  \country{Canada}
}

\author{Johannes Doerfert}
\email{jdoerfert@llnl.gov}
\affiliation{%
  \institution{Lawrence Livermore National Laboratory}
  \city{Livermore}
  \state{California}
  \country{USA}
}


\begin{abstract}
High-performance GPU programming has traditionally forced a compromise between execution efficiency and memory safety. While Rust guarantees compile-time memory safety for host CPUs via its strict ownership model, applying these constraints to massively parallel GPU execution environments has previously mandated either vendor-locked Domain-Specific Languages (DSLs) or escaping to explicit unsafe raw pointers. This paper presents a zero-overhead, multi-vendor GPU compilation framework built natively into the Rust compiler (rustc) and LLVM backends.

We leverage Rust’s rich type system, ownership system, and strict aliasing guarantees ($noalias$) to efficiently manage and optimize data transfers through LLVM’s Offload infrastructure.
We expose the technical challenges of cross-vendor ABI lowering mismatches between Host and Device targets and introduce a two-pass compilation pipeline capable of safely handling both manual and compiler-generated memory movements.
Evaluating our framework on RAJAPerf demonstrates that our rustc-based solution can generate competitive LLVM IR for GPU kernels, achieving a solid kernel performance against native, hand-optimized CUDA and HIP C++ baselines.
\end{abstract}

\maketitle


\input{sections/intro}

\begin{figure*}[t]
  \centering
  \begin{adjustbox}{max width=0.9\textwidth}
    \input{compilation_figure}
  \end{adjustbox}
  \vspace*{-2mm}
  \caption{Current three-pass implementation pipeline for Rust offloading.}
  \label{fig:offload-compile-flow}
\end{figure*}
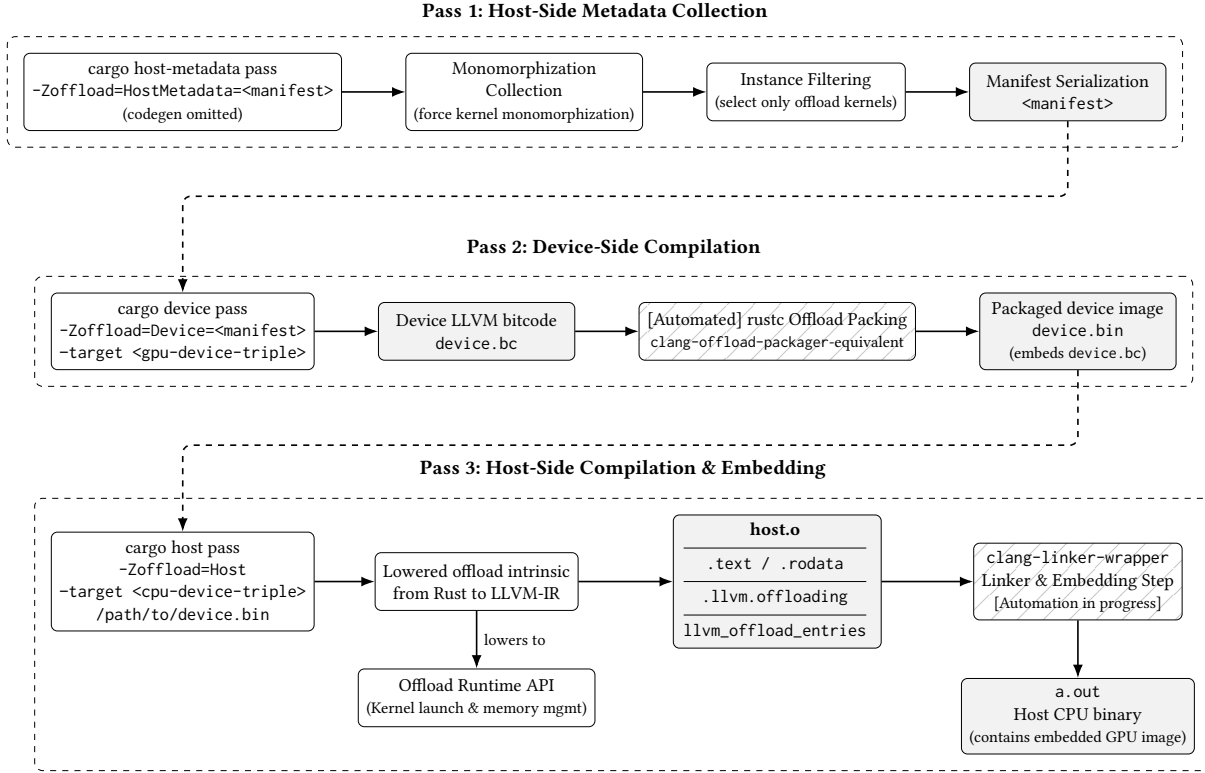

\input{sections/background}
\input{sections/safe_frontend}
\input{sections/toolchain}

\input{sections/optimizations}

\input{sections/evaluation}

\section{Related Work}
The Rust GPU ecosystem is under active development, with multiple approaches currently in progress:

\textbf{rust-gpu}: A SPIR-V-based approach\cite{RustGPU} which was originally targeting graphics programming in Rust, later extended to also target compute kernels.
Due to using the Vulkan flavour of SPIR-V, the rust-gpu project has to emulate pointers\cite{Rust-GPU}, which we consider a blocking issue for most HPC benchmarks.

\textbf{rust-cuda}: An NVIDIA-only approach\cite{RustCUDA} that uses raw pointers for all mutable arguments, consequently giving up on the portability and safety aspects our work provides.

\textbf{} A very early paper\cite{firstRustGPU} which precedes the first stable Rust release. It demonstrates GPU programming in Rust, by targeting both OpenCL and PTX directly. They also describe a two-pass compilation pipeline.

\textbf{cuda-oxide}: A new NVIDIA-led approach\cite{CudaOxide} to allow safe programming for NVIDIA GPUs. Although portability is not given, the safety ideas are comparable to those in our work and in \cite{drehwald2025taming}.
Beyond existing implementation differences, we are therefore curious if a unified frontend in the style of \cite{KAjl} is possible.
Their single-pass compilation makes a different trade-off from ours: it simplifies communication between host and device code, but frontend target-dependent decisions are evaluated only once. Our multi-pass design instead preserves independent host and device target semantics at the cost of explicit cross-pass communication.

Beyond Rust, our work shares commonalities with many of the common offloading languages.
A subset of CUDA was implemented on top of the OpenMP target infrastructure~\cite{CUDAviaOpenMP}, and we follow a similar path to reuse the OpenMP target efforts in LLVM~\cite{OpenMPGPUinClang}.
In LLVM, OpenMP target runtimes have been co-developed with (mostly) language specific compiler optimizations~\cite{EfficientOpenMPOnGPUs,codesignOpenMP}.
Similarly, we co-developed compiler optimizations for redundant data transfers, stemming from automatic data movement at kernel boundaries.
It is worth to note that while our optimizations are API specific, they apply to any language, which for now includes OpenMP target; a language that can suffer from the same issues if code is ported naively to the GPU.
Ongoing efforts are currently porting SYCL on top of the LLVM Offload infrastructure, which is the generic part spun out of OpenMP target.
As these efforts mature, we will port our approach to the more generic Offload APIs as well and enable our transformations for more input languages.

\section{Conclusions}

In this work, we presented a suite of portable interfaces for offloading Rust code to GPU accelerators, proving that memory safety does not preclude high performance. To cover a wide range of GPU programming paradigms, we provided three distinct interfaces: (1) offloading for standard Rust regions with automatic, compiler-managed data movement; (2) offloading via third-party libraries like cuBLAS and rocBLAS; and (3) a manual, user-controlled data movement interface. Regarding kernel programming, developers retain high flexibility: kernels can be implemented as unsafe code blocks, with CUDA and HIP-style thread indexing, or written entirely as safe code by leveraging our novel partitioning abstractions for transparent data-to-thread mapping. Built upon the LLVM Offload infrastructure, our interfaces achieve broad vendor portability, currently supporting NVIDIA and AMD accelerators, with planned support for Intel GPUs.

Our experiments demonstrate that this Rust offloading approach achieves near-parity with native kernel execution times on both NVIDIA and AMD GPUs. While we observed overheads related to total benchmark execution time---primarily stemming from current host-device synchronization inefficiencies---our approach stands as a highly competitive acceleration method for both safe and unsafe Rust code on major GPU vendors.


Future work will focus on minimizing device-host synchronization overhead by transparently integrating asynchronous data transfers and kernel launches. Furthermore, we plan to extend the frontend to support multi-device environments, enabling developers to scale their workloads across all available GPUs on a given node.

\section*{Acknowledgements}

We thank the Rust Foundation for their support of this work. Manuel Drehwald was supported in part by the Rust Foundation, and Marcelo Domínguez was supported in part through Google Summer of Code. The views and opinions expressed in this paper are those of the authors and do not necessarily reflect those of the Rust Foundation or Google.

We also especially thank Oli Scherer for discussion on supporting generics and for general code review, the Rust Infrastructure Team and Shota Sugano for their support in distributing our work through Rust's infrastructure, and Travis Cross for championing this work within the Rust Language Team.

Alán Aspuru-Guzik thanks Anders G. Fr{\o}seth for his generous support. Alán Aspuru-Guzik also acknowledges the generous support of Natural Resources Canada and the Canada 150 Research Chairs program.

This work was performed under the auspices of the U.S. Department of Energy by Lawrence Livermore National Laboratory under Contract DE-AC52-07NA27344 (LLNL-CONF-2023077).
This manuscript has been partially co-authored by Lawrence Livermore National Security, LLC under Contract No. DE-AC52-07NA27344 with the US. Department of Energy.
The United States Government retains, and the publisher, by accepting the article for publication, acknowledges that the United States Government retains a non-exclusive, paid-up, irrevocable, world-wide license to publish or reproduce the published form of this manuscript, or allow others to do so, for United States Government purposes.
\bibliographystyle{ACM-Reference-Format}
\bibliography{citations}


\end{document}

%% file: sections/intro.tex
\section{Introduction}

High-performance computing (HPC) and scientific applications remain heavily dominated by memory-unsafe languages like C, C++, or Fortran, with GPU support available through vendor-specific APIs, e.g., CUDA, and HIP, and portable programming models such as OpenMP, and SYCL~\cite{AnlSYCL}.
A defining characteristic of these paradigms is that they require developers to explicitly annotate data mappings, manually manage memory transfers, or embed their data into special buffers. 
Since these traditional programming languages lack the concept of safety boundaries, the burden of preventing memory corruption and data races falls entirely on the programmer.

Rust~\cite{RustLanguage} has emerged as a compelling alternative for modern systems programming, steadily gaining traction within the HPC~\cite{SafeRustHPC,ParallelRust,RustPerf} and Scientific Computing~\cite{NatureRust,Qiskit,Clarabel,RustBio} communities due to its ability to enforce\cite{RustBelt} memory safety and data-race freedom at compile time.

Furthermore, Rust's strict ownership and aliasing guarantees provide benefits that extend beyond basic software correctness; they provide such a good information baseline for the compiler backend, that we can start thinking about interfaces and optimizations that would barely be usable in unsafe languages. One of the examples is the  \texttt{noalias} metadata, which (almost) all safe Rust references receive by default, with no manual \texttt{restrict} annotations required. 

Despite these compiler advantages, utilizing Rust for heterogeneous acceleration remains bottlenecked by the lack of a portable GPU programming interface and a highly fragmented GPU software ecosystem. Existing GPU interfaces force developers into restrictive execution paradigms. For instance, \texttt{rust-gpu}, based on SPIR-V, has recently been re-oriented from graphics to compute workloads, but it lacks support for general pointers. Other compute-oriented solutions are either restricted to low-level bindings that require unsafe code regions for all kernels (\texttt{rust-cuda}), or are tightly coupled to specific hardware vendors (\cudaoxide). A portable interface and infrastructure that supports safe and idiomatic Rust across multiple vendors has so far been missing.

To close these gaps, we present a cross-vendor interface for GPU programming in Rust, integrated directly within the upstream Rust compiler (\texttt{rustc}), and based on the LLVM Offload infrastructure. 
This architecture generates native code for NVIDIA and AMD GPUs, and can extend to Intel and Apple targets as their upstream LLVM components mature. Our design introduces a two-pass compilation pipeline that strictly separates host and device intermediate representations (IR).

We designed three interfaces to accommodate a broad spectrum of use cases, ranging from managed, out-of-the-box offloading to explicit user-controlled offloading. The first interface allows developers to write native GPU kernels in safe and unsafe Rust with automatic data transfers and transparent optimizations. The second interface integrates vendor libraries like cuBLAS and rocBLAS, building upon these same automatic data-movement mechanisms. Lastly, the third interface permits manual execution of data transfers, granting developers absolute control over data movement, similar to CUDA or OpenMP.

Our key contributions are:
\begin{itemize}
    \item \textbf{Flexible GPU Programming:} Three offloading interfaces spanning from convenient wrappers for vendor libraries to explicit, type-enforced interfaces for custom kernel execution and manual memory management.
    \item \textbf{Safe memory access strategies:} A safe and extensible GPU programming frontend that separates parallel data-indexing and memory partitioning from the actual kernel execution, eliminating the need for user-written \texttt{unsafe} blocks in typical parallel workloads.
    \item \textbf{An In-Tree Cross-Vendor Toolchain:} Design and implement a portable two-pass compiler pipeline linking \texttt{rustc} to the LLVM Offload infrastructure, providing a unified target for NVIDIA and AMD hardware.
    \item \textbf{Type-Driven Offload Lowering:} A Mid-level Intermediate Representation (MIR) analysis enabling the automatic and efficient generation of memory transfers
    by leveraging Rust's type, layout, and mutability information, and the prevention of cpu to gpu communication bugs through the Ownership system. 
    \item \textbf{Multi-Vendor Benchmarking:} A performance evaluation using the RAJAPerf~\cite{RajaPerf} benchmark suite on AMD and NVIDIA accelerators. We demonstrate competitive GPU performance of the individual compute kernels.
\end{itemize}

%% file: compilation_figure.tex
\begin{tikzpicture}[
  >=Latex,
  font=\small,
  node distance=10mm and 11mm,
  process/.style={
    draw,
    rectangle,
    rounded corners=3pt,
    align=center,
    minimum width=34mm,
    minimum height=10mm,
    inner sep=4pt
  },
  artifact/.style={
    draw,
    rounded corners=3pt,
    align=center,
    minimum width=34mm,
    minimum height=10mm,
    inner sep=4pt,
    fill=gray!10
  },
  automated/.style={
    process,
    pattern={
      Lines[
        angle=45,
        distance=7pt,
        line width=0.35pt
      ]
    },
    pattern color=gray!45
  },
  group/.style={
    draw,
    rounded corners,
    dashed,
    inner sep=8pt
  },
  arrow/.style={->, thick}
]

\node[process] (rustc1) {
  cargo host-metadata pass\\
  \texttt{-Zoffload=HostMetadata=<manifest>}\\
  {\footnotesize (codegen omitted)}
};

\node[process, right=of rustc1] (mono) {
  Monomorphization\\
  Collection\\
  {\footnotesize (force kernel monomorphization)}
};

\node[process, right=of mono] (filter) {
  Instance Filtering\\
  {\footnotesize (select only offload kernels)}
};

\node[artifact, right=of filter] (manifest) {
  Manifest Serialization\\
  \texttt{<manifest>}
};

\draw[arrow] (rustc1) -- (mono);
\draw[arrow] (mono) -- (filter);
\draw[arrow] (filter) -- (manifest);

\node[process, below=28mm of rustc1] (rustc2) {
  cargo device pass\\
  \texttt{-Zoffload=Device=<manifest>}\\
  \texttt{--target <gpu-device-triple>}
};

\node[artifact, right=of rustc2] (devbc) {
  Device LLVM bitcode\\
  \texttt{device.bc}
};

\node[automated, right=of devbc] (packager) {
  [Automated] rustc Offload Packing\\
  {\footnotesize \texttt{clang-offload-packager}-equivalent}
};

\node[artifact, right=of packager] (hostout) {
  Packaged device image\\
  \texttt{device.bin}\\
  {\footnotesize (embeds \texttt{device.bc})}
};

\draw[arrow] (rustc2) -- (devbc);
\draw[arrow] (devbc) -- (packager);
\draw[arrow] (packager) -- (hostout);

\node[process, below=28mm of rustc2] (rustc3) {
  cargo host pass\\
  \texttt{-Zoffload=Host}\\
  \texttt{--target <cpu-device-triple>}\\
  \texttt{/path/to/device.bin}
};

\node[process] (lowering) at (devbc |- rustc3) {
  Lowered offload intrinsic\\
  from Rust to LLVM-IR
};

\node[process, below=10mm of lowering] (runtime) {
  Offload Runtime API\\
  {\footnotesize (Kernel launch \& memory mgmt)}
};

\node[artifact, rectangle, rounded corners=3pt, inner sep=0pt, minimum width=36mm] (hosto) at (packager |- rustc3) {
  \begin{tabular}{@{}c@{}}
    \rule{0pt}{3.5mm}\textbf{host.o}\\[1mm]\hline
    \rule{0pt}{3.5mm}\texttt{.text / .rodata}\\[1mm]\hline
    \rule{0pt}{3.5mm}\texttt{.llvm.offloading}\\[1mm]\hline
    \rule{0pt}{3.5mm}\texttt{llvm\_offload\_entries}\\[1mm]
  \end{tabular}
};

\node[automated] (linker) at (hostout |- rustc3) {
  \texttt{clang-linker-wrapper}\\
  Linker \& Embedding Step\\
  {\footnotesize [Automation in progress]}
};

\node[artifact, below=10mm of linker] (finalbin) {
  \texttt{a.out}\\
  Host CPU binary\\
  {\footnotesize (contains embedded GPU image)}
};

\draw[arrow] (rustc3) -- (lowering);
\draw[arrow] (lowering) -- (hosto);
\draw[arrow] (hosto) -- (linker);
\draw[arrow] (linker) -- (finalbin);
\draw[arrow] (lowering) -- node[midway, right, font=\footnotesize] {lowers to} (runtime);

\coordinate (pass1_drop) at ($(manifest.south)+(0,-13mm)$);
\coordinate (pass2_drop) at ($(hostout.south)+(0,-13mm)$);

\draw[arrow, dashed, rounded corners=4pt] 
  (manifest.south) -- (pass1_drop)
  -- (rustc2.north |- pass1_drop) -- (rustc2.north);

\draw[arrow, dashed, rounded corners=4pt] 
  (hostout.south) -- (pass2_drop)
  -- (rustc3.north |- pass2_drop) -- (rustc3.north);

\node[group, fit=(rustc1)(mono)(manifest)] (group1) {};
\node[font=\bfseries, above=2mm of group1.north] 
  {Pass 1: Host-Side Metadata Collection};

\node[group, fit=(rustc2)(devbc)(packager)(hostout)] (group2) {};
\node[font=\bfseries, above=2mm of group2.north] 
  {Pass 2: Device-Side Compilation};

\node[group, fit=(rustc3)(lowering)(runtime)(hosto)(linker)(finalbin)] (group3) {};
\node[font=\bfseries, above=2mm of group3.north] 
  {Pass 3: Host-Side Compilation \& Embedding};

\end{tikzpicture}

%% file: sections/background.tex
\section{Background}

\subsection{Rust}
\label{sec:background-rust}
While a complete introduction is out of scope, we will give a very brief summary of the aspects of Rust which we consider as most valuable to understand this paper.

\paragraph*{\textbf{Unsafe}}
A strength of Rust is the explicit separation of safe and unsafe Rust. It is a common goal to isolate unsafe operations, and then provide safe abstractions on top. We ``know'' that GPU programming is not inherently unsafe - Rust currently just often leaves us without an opportunity to express that our parallel writes into an object are safe.
In this work, we explore how such safe abstractions for GPU programming might look like.

\paragraph*{\textbf{References and Raw Pointers}}
Rust offers raw, c-style pointers, e.g. \lstinline{*mut f32}. Their usage is very uncommon outside of Foreign-Function-Interfaces (FFI). References provide various benefits over raw pointers. The compiler enforces that references can not be used after the object to which they point went out of scope or got dropped. As such, there is no null-pointer equivalent for references. There are two types of references: const and unique (aka. \lstinline{mut}able). If a unique reference (\lstinline{&mut T}) exists, no other reference (neither const nor mut) may exist to the same object. However, if no mutable reference exists, arbitrary many const references (\lstinline{&T}) may exist. This rule is often referred to as the \textit{aliasing XOR mutability rule}. More detailed models are explained in \cite{TreeBorrows} and \cite{StackedBorrows}. A user can only create a mutable reference if the underlying object was defined as mutable. 
\begin{lstlisting}[language=Rust, style=boxed,basicstyle=\small]
let mut x: Vec<i32> = vec![10; 1024];
let y: &mut [i32] = &mut x;
\end{lstlisting}

The compiler enforces all of these requirements. On the LLVM-Intermediate Representation (LLVM-IR) level, references are represented as pointers with the \lstinline{noalias} attribute. The only exception is described next.

\paragraph*{Interior Mutability}
\label{sec:unsafe_cell}
An advanced concept which allows users to create a const reference to a mutable object wrapped in an \lstinline{UnsafeCell}. For example, a reference of type \\
\lstinline{&UnsafeCell<i32>} can be used to mutate an underlying \lstinline{i32} value, even if the reference is not unique. 

\paragraph*{\textbf{Generics}}
Generic functions and structs in Rust are similar to templates in C++. They are instantiated, aka ``monomorphized'', at compile time, based on the concrete types with which they are invoked.

\paragraph*{\textbf{Lifetimes}}
All references have a lifetime, which ties them to their underlying object. Lifetimes can usually be inferred by the compiler, but at times have to be annotated explicitly with an apostrophe: \lstinline{&'a f32}. An example from the Rust docs\cite{Lifetimes} shows an application:
\begin{lstlisting}[language=Rust, style=boxed,basicstyle=\small]
fn longest<'a>(x: &'a str, y: &'a str) -> &'a str {
    if x.len() > y.len() { x } else { y }
}
\end{lstlisting}
This function might return either of the two references. As such, the returned reference is only alive while both of the input references are alive. They therefore use the same lifetime.


\paragraph*{\textbf{config (``cfg'')}} Configurations in packages serve two main purposes. Compile times in Rust are a major concern, so most crates offer a larger number of features.
A crate user can then select in their \lstinline{Cargo.toml} which features of which dependency they want to enable or disable.
Inside their crate, a developer can use the \lstinline{#[cfg(CONFIG == VALUE)]} syntax to ask the Rust compiler if a certain config was set to the given value.
If a \lstinline{cfg} attribute evaluates to false, the guarded function, module, or struct will be discarded at the Abstract Syntax Tree (AST) level, to minimize compile-time impact. Compute heavy crates~\cite{faer} often also use different \lstinline{cfg} checks to provide multiple algorithm implementations that are optimized for different compilation targets like aarch64 or x86\_64 and use intrinsics that are not generally available.

\subsection{LLVM/OpenMP and Libomptarget}
We build our work on top of LLVM's OpenMP/Offload library. This has the benefit, that we are able to support both the ``nvptx64-nvidia-cuda'' and ``amdgcn-amd-amdhsa'' target, which are already available in today's Rust compiler. An Intel GPU target is under development, which will allow us to support safe abstractions for all major GPU vendors.

\paragraph*{\textbf{OpenMP target vs. Offload}}
OpenMP supports parallel CPU and GPU programming. LLVM originally implemented the required OpenMP GPU support within the OpenMP target runtime. Later, GPU Offload was split from OpenMP to make it easier for other language frontends to utilize the functionality. This refactoring is still in progress. Rustc currently generates OpenMP target runtime calls, but we consider this an implementation detail and will transfer to Offload calls in the near future. For consistency, we will generally refer to Offload, unless talking about concrete elements like the libomptarget library.

%% file: sections/safe_frontend.tex
\section{A Frontend For Safe GPU Programming}\label{sec:frontend}

\subsection{Heterogeneous Execution Models}\label{subsec:exec_models}
We define three programming models, from a fully automated host-managed execution to explicit, type-enforced device memory layout control.

\subsubsection{Interface A: Compiler-Managed Rust Kernels}
Interface A is the default interface for writing GPU kernels in Rust. The
programmer writes a Rust function over ordinary references and invokes it with
an offload macro:

\begin{lstlisting}[language=Rust, style=boxed, basicstyle=\small\ttfamily]
offload!(matrix_multiply, 
  &matrix_a, &matrix_b, &mut matrix_c
);
\end{lstlisting}

The callee is compiled by rustc for the device target. The
host invocation is lowered into runtime calls that allocate device
storage, transfer inputs, launch the generated kernel, and synchronize outputs
according to the Rust types at the call boundary. Immutable references
(\texttt{\&T}) are treated as read-only inputs, while mutable references
(\texttt{\&mut T}) are values that may be written by the kernel and
therefore must be made visible on the host after the call.

This interface is intentionally plug-and-play: callers do not explicitly
manage device memory. That convenience also defines its performance limit.
Because each offload operation is specified in terms of host references, an
intervening host use of a value forces the runtime to make the latest device
value visible on the CPU. Consequently, chains of kernels can suffer from
implicit host-device synchronization unless the compiler can prove that a
transfer is redundant.

\begin{lstlisting}[language=Rust, style=boxed,basicstyle=\small\ttfamily]
offload!(kernel_1, &input, &mut output);
// Host usage of output forces a datatransfer
println!("{:?}", &output);
offload!(kernel_2, &input, &mut output);
\end{lstlisting}

Although the program does not mention memory movement, the host read of
\texttt{output} forces a device-to-host synchronization after
\texttt{kernel\_1}. The second kernel then requires the updated value to be
available on the device again, inducing a host-to-device transfer before
\texttt{kernel\_2}. This behaviour is safe and convenient, but it makes data
movement depend on seemingly unrelated host-side uses such as logging or
debugging.

\subsubsection{Interface B: Interoperability with Host-launched GPU Vendor Libraries}
Many GPU applications obtain their performance from highly
optimized vendor libraries such as cuBLAS or rocBLAS. These are
called from host code and internally manage their own kernel launches, so
the Rust compiler cannot optimize or inspect the device code itself. However,
the compiler can still use the Rust type system to reason about the data passed across the call boundary. 

Interface B treats such calls as host-side offload operations.
Our intrinsic defaults to automatic data transfers to and from the device before and after invoking the underlying library, but other behaviour like forwarding existing GPU pointers can also be specified.
The compiler uses the same mapping infrastructure as for GPU kernels to materialize the required device pointers before invoking the library API.
\begin{lstlisting}[language=Rust, style=boxed, basicstyle=\small\ttfamily]
core::intrinsics::offload_args::<_, _, ()>(
  rocblas_sgemv_wrapper, 
  (&A, &x, &mut y)
);
\end{lstlisting}
The main benefit is that with this intrinsic normal Rust functions, vendor library calls and Rust GPU kernels can share one interface.
This lets programmers incrementally replace hot CPU operations with GPU vendor implementations. 
It also allows the compiler to hoist,
reuse, or eliminate data transfers around the call boundary. We describe a set of optimizations to leverage this opportunity in \ref{sec:optimizations}.
It also gives the runtime a uniform point at which to collect profiling data and diagnostics for offloaded operations, even when vendor-library internals remain opaque.

\subsubsection{Interface C: Explicit Type-Staged Memory Control}

Interface C makes the Host or Device location of data explicit through Rust types. In Interface B,
host-side uses can silently introduce transfers between kernels. Interface C
instead exposes this synchronization point in the program: values that may be
modified on the device are managed through a type that keeps the corresponding
host object borrowed until the GPU version of the value is dropped.


\begin{lstlisting}[language=Rust, style=boxed,basicstyle=\small\ttfamily]
pub struct Preload<'a, T: ?Sized> {
  cpu_ptr: *const T,
  _marker: PhantomData<&'a T>,
}
pub struct PreloadMut<'a, T: ?Sized> {
  cpu_ptr: *mut T,
  _marker: PhantomData<&'a mut T>,
}
\end{lstlisting}

These types conservatively store raw host pointers rather than Rust references.
A Rust reference has significantly higher correctness requirements than a raw pointer. For example the referenced memory
must be valid for the duration of the borrow and the usual aliasing rules must be upheld whenever the reference is used. 
After preloading, however, the most recent
version of the data will live on the device, whose allocation has a
Different address from the original host allocation. The host pointer is
therefore only a runtime key identifying the mapped allocation, not a reference
to the current contents of the value.

While we intentionally avoided references, the preloaded value must still participate in Rust's lifetime and aliasing rules, to avoid trivial Undefined Behaviour (UB).
The \texttt{PhantomData} field provides this connection
without requiring the struct to contain an actual reference. A
\texttt{Preload<'a, T>} carries the lifetime of an immutable borrow, while a
\texttt{PreloadMut<'a, T>} carries the lifetime of a mutable borrow. 
The two variants express different access guarantees. First, \texttt{Preload<'a, T>}
is read-only: it behaves like an immutable borrow of the host value, so other
host reads remain valid, but mutation is forbidden while the Preload handle
exists. Second, \texttt{PreloadMut<'a, T>} may be modified by device code: it behaves
like a mutable borrow, so Rust code cannot read or write the original host value
until the GPU handle is dropped. This follows Rust's standard aliasing rule: either one mutable reference exists, or any number of immutable
references exist, but not both.

\begin{lstlisting}[language=Rust, style=boxed,basicstyle=\small\ttfamily]
fn main() {
  let mut output = vec![0.0f32; 1024];

  // Stage output on the device; output is
  // borrowed until out_gpu is dropped.
  let out_gpu =
      core::intrinsics::preload_mut(
          &mut output
      );

  offload!(kernel_1, &input, &out_gpu);
  offload!(kernel_2, &input, &out_gpu);

  // Explicitly synchronize back to host.
  drop(out_gpu);

  println!("{:?}", output);
}
\end{lstlisting}

The LLVM backend of rustc lowers the constructor of the Preload values into an offload begin-data-mapper operation and lowers drop ("destructor") calls into the corresponding end-mapper operation.
Calling preload multiple times on a immutable reference simply creates multiple Preload handles and increases a reference counter in the runtime.
Dropping a Preload Type therefore decreases the reference counter and only results in an actual freeing of the gpu memory location if the counter reaches zero.
PreloadMut types can not alias by design, and as such the reference counter is irrelevant. Each drop call does not only result in a free, but also generates a Device to Host transfer, since the underlying GPU allocation has likely been modified.
The memory transfer at the begin of a Kernel 
launch therefore becomes a no-op, if all arguments have been preloaded.

This design trades some convenience for predictable performance. An intermediate
host read of \texttt{output} can no longer silently introduce a transfer between
\texttt{kernel\_1} and \texttt{kernel\_2}. With \texttt{PreloadMut}, such a read
is rejected by the borrow checker until the mutably preloaded value is dropped; the drop
marks the explicit point at which the device result is synchronized back to the
host. Interface C is therefore the appropriate interface for more complex gpu pipelines and library APIs that expect inputs or outputs to already be present on the GPU.

\subsection{Safe Rust Kernels}
\label{subsec:region_partition}
As described in Section~\ref{sec:background-rust}, references in Rust follow the \textit{aliasing XOR mutability rule}. A mutable reference therefore might be better described as a unique reference. This is clearly at odds with even the most basic \texttt{vec\_add} example:
\begin{lstlisting}[language=Rust, style=boxed,basicstyle=\small\ttfamily]
fn vec_add(a: &[f32], b: &[f32], c: &mut [f32]) {
  let idx = _block_dim_x();
  c[idx] = a[idx] + b[idx];
}
\end{lstlisting}
In GPU programming, kernels are often executed with multiple threads, which can cause UB when sharing mutable references. The most direct solution is turning all mutable inputs into raw pointers, as required by \texttt{rust-cuda}. However, building a GPU ecosystem on unsafe code blocks diminishes the purpose of the Rust language.
Drehwald et al.~\cite{drehwald2025taming} and \cudaoxide introduce a path towards safe GPU programming by decoupling the memory access patterns from kernel implementation. The underlying idea is that slices should fundamentally work safely, given that in most kernels, threads access disjoint elements of the slices they are modifying. The challenge therefore lied in expressing this safely within the language, to the compiler.

We want users to write safe Rust using standard slices, so our frontend handles it with raw pointers under the hood.
We introduce an abstraction called a \verb|Region|, which lets users safely work with slices in a parallel environment.

\begin{lstlisting}[language=Rust, style=boxed, basicstyle=\small\ttfamily]
let mut x = [0.0f64; 256];
let mut reg =
    Region::<_, Linear1D>::new(&mut x, ());
\end{lstlisting}

Every \verb|Region| is tied to a \verb|PartitioningStrategy|. This strategy decides exactly which elements each thread is allowed to read and write to, ensuring that no pair of threads ever get overlapping memory regions.

\begin{lstlisting}[language=Rust, style=boxed,basicstyle=\small\ttfamily]
pub unsafe trait PartitioningStrategy {
  type Shape: Copy;
  type View<'a, T: 'a>;
  type ViewMut<'a, T: 'a>;

  unsafe fn get<'a, T>(
      ptr: *const T,
      len: usize,
      shape: Self::Shape,
  ) -> Option<Self::View<'a, T>>;
  unsafe fn get_mut<'a, T>(
      ptr: *mut T,
      len: usize,
      shape: Self::Shape,
  ) -> Option<Self::ViewMut<'a, T>>;
}
\end{lstlisting}
The trait here is unsafe, because any incorrect implementation of a PartitionStrategy will likely result in Undefined Behaviour. Beyond that, both the \texttt{get} and \texttt{get\_mut} functions are unsafe, since the passed pointer must point to an allocation with at least \texttt{len} elements. Neither of these requirements can be verified by the compiler, therefore both trait and functions are unsafe.

The safety is introduced by the individual implementations of the PartitionStrategy, which promise to uphold these invariants. To support our benchmarks in \ref{sec:benchmarks}, we implemented multiple PartitionStrategy variants. We want to emphasize that all these strategies are implemented in pure Rust, without relying on any internal compiler features or any compiler internal abilities. While it seems sensible to place at least the trait definition and some of the more popular strategies into the standard library, we have no technical need to do so. By releasing our interface as a standalone crate, we hope to encourage users to explore additional partitioning schemes.

\paragraph{Comparison with \cudaoxide}
While our implementation is highly inspired by the design of Drehwald et al.~\cite{drehwald2025taming}, we want to analyze the differences with the \cudaoxide frontend. Although \cudaoxide has a similar design, there are three primary differences: Firstly, their \texttt{DisjointSlice} is part of the \cudaoxide project and can not be extended by users. Secondly, their design is currently scalar oriented, whereas our interface allows the return of disjoint chunks of data. Finally, \cudaoxide allows safe indexing into their \texttt{DisjointSlice} via a special \texttt{ThreadIndex} object with an advanced set of type system checks, while our PartitionStrategies skip this complexity, compute the index internally, and directly return a mutable reference for each thread.

While both frontends currently support a different set of access patterns, we see no fundamental reason why they can not be extended to support the same kernels. We hope that through enough end-user feedback \texttt{Rust offload} and \cudaoxide can eventually converge into a single design for safe kernels. Outside of the frontend, both projects diverge on other major axes (vendor-agnostic vs. vendor specific, single-pass vs. multi-pass compilation). If both projects can use a similar or even identical frontend, we hope that the remaining Rust GPU projects will also adopt a compatible design to achieve safety.



\subsection{Exposing GPU Shared Memory}\label{subsec:shared_mem}
Advanced GPU algorithms frequently utilize shared memory as a high-performance scratchpad~\cite{ButterflySharedmem} to store temporary variables within kernels.
This memory space is private to each thread block, allowing threads within the same block to efficiently exchange data.
In Section \ref{subsec:region_partition}, we described how we can safely encode strictly disjoint thread-access patterns to cover most use cases for input arguments by assigning disjoint subsections to individual threads.
However, shared memory usually has the opposite purpose, as threads actively cooperate over this block-local storage to execute efficient block-level primitives such as reductions and tiled operations.
To demonstrate its use in Rust, in \ref{sharedmem} we show how a kernel can requests two blocks of $\lstinline{BLOCK_SIZE}^2$ bytes from the shared memory space, in order to perform a tiled matrix multiplications within each block:

\begin{lstlisting}[language=Rust, style=boxed,basicstyle=\small\ttfamily]
core::intrinsics::offload::<_, _, ()>(
    gpu_square_matrix_mult,
    [grid_cols, grid_rows, 1],
    [nthreads, nthreads, 1],
    (2 * BLOCK_SIZE * BLOCK_SIZE) as u32,
    ..args
)
\end{lstlisting}

\begin{lstlisting}[language=Rust, style=boxed,basicstyle=\small\ttfamily]
pub unsafe extern "gpu-kernel" 
fn gpu_square_matrix_mult(..args) {
unsafe {
  const LENGTH: usize =
    (BLOCK_SIZE * BLOCK_SIZE) as usize;
  let tile_a =
    gpu_launch_sized_workgroup_mem::<i32>()
      as *mut [i32; LENGTH];
  let tile_b =
    gpu_launch_sized_workgroup_mem::<i32>()
      .add(len) as *mut [i32; LENGTH];
  ...
}
\end{lstlisting}\label{sharedmem}

As we can see in the example, exposing raw accelerator shared memory to high-level code introduces four distinct safety hazards:
\begin{enumerate}
    \item Accessing an object stored on shared memory whose size exceeds the shared memory bytes requested at kernel launch can lead to out-of-bounds accesses.
    \item If the shared memory space holds multiple objects, applying an incorrect offset when accessing a specific object can lead to incorrect outcomes and memory errors.
    \item Casting a shared memory pointer to a standard Rust reference (\texttt{\&T} or \texttt{\&mut T}), while multiple threads concurrently read and write from the memory at that address, causes UB.
    \item The guaranteed alignment of the pointer returned by \texttt{gpu\_launch\_sized\_\-workgroup\_\-mem::<T>} is the alignment of type \texttt{T}. This can lead to UB if the pointer is later casted to a type with a higher alignment requirement.
\end{enumerate}


\noindent
Returning a raw pointer in \texttt{gpu\_launch\_sized\_\-workgroup\_\-mem::<T>} is already a clear signal to developers. Since the pointer is only usable within device code, additional safety guardrails seem unlikely to justify their complexity.

%% file: sections/toolchain.tex
\section{Toolchain \& Lowering Implementation}

The Rust compiler supports two modes of compilation. Code can either be compiled for the host on which the compiler is running or cross-compiled for a different target. We need both abilities when compiling a single codebase for the CPU host and the GPU target. Existing compiler infrastructures typically address this using either a single-pass or a two-pass compilation pipeline. In a single-pass pipeline, the compiler frontend (e.g., \texttt{rustc}) is invoked once, duplicating an intermediate representation (IR) to target each architecture. For our toolchain, we implement a two-pass compilation pipeline, as shown in Figure~\ref{fig:offload-compile-flow}.

To give a realistic pipeline overview, we demonstrate our pipeline using Rust's official build manager, \texttt{cargo}, even though our implementation modifications reside entirely within the compiler frontend (\texttt{rustc}) and the LLVM backend. The experimental \texttt{-Z offload=Device} flag is forwarded transparently from \texttt{cargo} to \texttt{rustc}, requiring no changes to the build manager itself and ensuring that alternative build systems like Bazel or Buck2 can support our offloading workflow with minimal changes. This device pass uses existing upstream target definitions for AMD and NVIDIA architectures (with ongoing development for Intel GPU support). Our modified pipeline only alters the final stage of code generation: instead of just emitting device bitcode, our compiler wraps the bitcode into a device binary. This part is similar to the Clang-based C++ OpenMP toolchain, although Clang delegates this packaging step to an external \texttt{clang-offload-packager} executable, while rustc invokes the underlying Offload APIs directly. This allows us to minimize the number of binaries involved in our pipeline.

The second \texttt{cargo} invocation manages the host-side compilation. We again trigger our modified pipeline via the modified pipeline flag \texttt{-Z Offload=Host=/path/to/device.bin}. During this pass, the compiler lowers our new Rust offload intrinsics introduced in Section~\ref{sec:frontend} into OpenMP target (i.e., \texttt{libomptarget}) runtime calls.
Following the intrinsic lowering, our toolchain embeds the device binary generated in the previous pass directly into the host LLVM IR to produce the final host object. Notably, because the underlying LLVM Offload infrastructure supports fat binaries, this embedded payload can bundle multiple device binaries, which allows support for both AMD and NVIDIA architectures from one executable. To finalize the compilation, the pipeline invokes the \texttt{clang-linker-wrapper} executable, which links the required offload runtime libraries into the executable.

While it may appear cumbersome to pass the device binary path explicitly between compiler invocations, it is intentional. Even though \texttt{rustc} and \texttt{cargo} are co-developed within the same GitHub organization, rustc is prohibited from making assumptions regarding the filesystem location of intermediate build artifacts. This separation for example simplifies caching, and generally just ensures that \texttt{cargo} and \texttt{rustc} can be developed and improved independently. End-users do not need to run this multi-step sequence manually, since the entire workflow can be handled by a standard \texttt{cargo} subcommand wrapper.

\paragraph{Alternative Design Considerations:} An alternative design which we considered was a single-pass implementation, in which we would only invoke cargo once, similar to \cudaoxide. 
The Rust compiler fundamentally expects one compilation target per invocation. If that target is also used to determine both host and device target semantics, we would need to choose between the CPU target \\(e.g. \texttt{x86\_64-unknown-linux-gnu}), the GPU target \\(e.g. \texttt{amdgcn-amd-amdhsa}), or a new combined target. None of these choices seemed optimal to us. An alternative, as used by cuda-oxide, is to retain the CPU target for frontend compilation and separate device code at a later stage. This preserves host-target semantics, but frontend target-dependent decisions such as \texttt{cfg} evaluation are still made only once.
Rust is a systems language that has direct support for both inline assembly and target specific intrinsics like avx512 or neon instructions. Compute-heavy crates can therefore provide a default implementation, supported by faster but vendor-specific implementations which are gated behind \texttt{\#[cfg(target\_arch=<X>)]} checks. 
Using the CPU target preserves these host-side optimizations, but it means that \texttt{cfg} evaluation may select CPU-specific implementations before host and device code are separated. Consequently, MIR reachable from a GPU kernel can contain CPU-specific intrinsics or inline assembly.
For the general case, we consider translating such target-specific code into GPU-compatible IR infeasible. We therefore favour separate frontend passes, which allow target-dependent source constructs to be evaluated independently for the CPU and GPU.
This choice comes with a cost: single-pass compilation keeps host and device information within one compilation, simplifying cross-target communication and monomorphization, while our multi-pass design must explicitly communicate this information between passes.

It might seem tempting to compromise between the two approaches. A granular compromise would require us not to evaluate \texttt{cfg} attributes and macros in the AST, and instead carry them throughout the compiler, till our final MIR layer, where we would split the IR between host and device IR. This would require adjusting most IR and lowering code, and furthermore pessimize compile times and memory usage. To avoid imposing this overhead on non-GPU users, rustc would need to retain its existing behaviour for ordinary compilations, adding further implementation complexity.
A more coarse compromise would be to simply have two completely independent AST and IR copies in a single rustc invocation. 
While the coarse compromise would simplify some of our communication, we believe the actual simplifications for us would not outweigh the increased complexity in other parts of the compiler. For now, we therefore use a two-pass compilation also performed by OpenMP offload, HIP, CUDA, and others.

\subsection{Lowering of our Offload intrinsics}\label{subsec:lowering}

Discrete accelerator memory spaces typically require explicit, developer-directed data mapping, as alternative Unified Shared Memory (USM) abstractions or software-managed runtimes often introduce notable performance overhead and microarchitectural constraints~\cite{AnlSYCL}. Rather than forcing programmers to manually annotate data-sharing boundaries via verbose pragmas, as is common in OpenMP or SYCL, our toolchain derives data directionality semantics automatically from Rust's type system and ownership model.

Because safe Rust enforces immutability by default and triggers compiler warnings for redundant \texttt{mut} qualifiers, argument mutability provides a sound signal for data directionality. During Rust MIR lowering, the framework walks the kernel parameter layouts to compute the number of bytes to be transferred and to synthesize corresponding LLVM \texttt{Libomptarget} data-mapping clauses:
\begin{itemize}
    \item Immutable references (\texttt{\&T}) and constant raw pointers (\texttt{*const T}) lower to the \texttt{MapTo} directive, copying data exclusively to the device.
    \item Mutable references (\texttt{\&mut T}) and mutable raw pointers (\texttt{*mut T}) lower to the bidirectional \texttt{MapToFrom} directive and enable post-kernel synchronization.
    \item Scalar arguments up to 64 bits are passed by value using OpenMP target's \texttt{IMPLICIT} and \texttt{LITERAL} flags, bypassing pointer indirection and allocation overhead.
\end{itemize}


\paragraph{Cross-Boundary Monomorphization}\label{par:generics}
Idiomatic Rust relies on compile-time monomorphization to generate specialized, concrete function instances from generic definitions. This mechanism is important for GPU targets, where static type resolution enables additional LLVM optimizations. 
However, our two-pass compilation pipeline breaks standard monomorphization tracking.
Because the host application's entry point (\texttt{main}) is omitted or conditionally compiled out during the device pass, the concrete type substitutions requested by host-side \texttt{offload!} invocations are invisible during the device compilation pass.

To resolve this mismatch, we considered two strategies:

\begin{enumerate}
    \item \textbf{Dual-Pass Main Compilation:} Compiling the host entry point (main) during the device pass solely to force kernel instantiations. This approach has a similar challenge to a single-pass solution; target-specific conditional compilation (\texttt{\#[cfg]}) can cause the device pass to miss code paths that are active during the host pass.
    \item \textbf{Cross-Pass Metadata Exportation:} Leveraging a dedicated compiler pass to serialize kernel definition identifiers (\texttt{DefId}) and their concrete type substitutions during the host phase. The subsequent device pass imports this metadata to seed its root monomorphization collection. 
\end{enumerate}

We implemented the second approach as a new query in the Rust compiler. The compilation overhead of this metadata serialization is expected to be negligible, as it integrates with incremental compilation caching, and the first host pass can skip the expensive codegen part.

\subsection{Discussion on Future Support}\label{subsec:impl_future_work}

While we implemented most of our desired features in this prototype, there are two important features that we plan to add or extend, namely the support to pass additional types between the host and device, and the ability to run most of the standard library on GPUs.

\paragraph{Frontend ABI Validation}\label{subsec:abi_val}
Since our two-pass architecture decouples host and device compilations, type layouts and ABI lowerings can diverge between the two targets.
For instance, we identified a target discrepancy in how primitive slices are lowered: the \texttt{x86\_64} host and \texttt{amdgcn} backends represent a slice as two scalar values (\texttt{ptr, int}), whereas the upstream \texttt{nvptx64} target lowers it as a fixed-size array (\texttt{[i64; 2]}).
While we are engaging with upstream maintainers to unify these lowering targets where sensible, the divergence, even on the basic slice type, shows the necessity of automated cross-boundary validation before adding support for more advanced types like structs.

This static validation is only required for host-device kernel entry points; GPU-to-GPU device calls are lowered by the same target, so arbitrary types can be passed.

\paragraph{Standard Library Support on GPUs}\label{subsec:std_gpu}

Providing a full GPU implementation of Rust's \texttt{std} is outside the scope of this work. 
Reimplementing broad standard-library support inside the Rust frontend would also duplicate a difficult and largely orthogonal engineering effort.
We instead plan to adopt the idea established by LLVM's \texttt{libc-for-gpu}~\cite{LIBCforGPUsPaper,huberLIBCforGPU} and later replicated for Rust (outside of LLVM) in \cite{VectorWareStd}.

%% file: sections/optimizations.tex
\section{Compiler Optimization Passes}\label{sec:optimizations}

Rust Offload provides two interfaces which support Rust-written GPU kernels: an explicit interface, where users control data movement directly, and a more convenient interface, where data is transferred automatically for each kernel launch. The convenient interface is easier to use, but it can be substantially slower when the same data is reused across multiple kernels. Without relying on the evaluation results in detail, our experiments show that repeated automatic transfers can make this interface over 400$\times$ slower than the explicit version with dedicated data movement.

We expect that most Rust offload users will start prototyping GPU applications with our convenient interface.
We therefore propose a set of optimizations that, for the common case, can make the convenient interface as performant as manual transfers.
These optimizations would be enough to match the overall runtime of our explicit interface on the evaluated RAJAPerf Benchmarks. 
We prototyped most of these optimizations as an extension to the LLVM OpenMP-opt pass~\cite{codesignOpenMP,EfficientOpenMPOnGPUs}.
Based on initial experiments, we are confident that with further testing and upstreaming, we can eliminate the need for most Rust Offload users to reach for the dedicated data movement interface.

\paragraph{Automatic data prefetching}
LLVM's offload API provide an interface which already matches our convenient offload intrinsic. 
In a single call, it transfers all arguments to the GPU, launches a kernel, and transfers some arguments back to the host.
This is not easy to optimize, so we adjusted our code generation to be more explicit.
Each offload intrinsic now generates three calls: (1) a Host2Device transfer for all arguments, (2) a kernel launch, (3) a Device2Host transfer for selected arguments.
As part of this transfer, we prototyped an extension to turn these blocking transfers into async ones.
The Host2Device is started earlier, if we can prove that it is legal to do so.
We then await it at the kernel launch location.
Similarly, the Device2Host transfer is started directly after the kernel has terminated, but we only block to await it before its next use~\cite{AsyncDataTransfers}.

\paragraph{LICM}
The second optimization we prototyped is a variant of Loop-Invariant-Code-Motion (LICM).

\begin{lstlisting}[language=Rust, style=boxed,basicstyle=\small]
for i in 0..100 {
  offload!(vec_add, &a, &b, &mut c);
}
\end{lstlisting}

Based on our work in the previous step, we have already generated separate memory transfer and kernel launch instructions.
We extended our previous optimization to try and hoist both transfers out of the loop.
If we succeed, we could then continue to turn them into async transfers for further improvements.

In order to also handle unrolled loops, we prototyped handling of repeated kernel calls, which might only differ by scalar offsets:
\begin{lstlisting}[language=Rust, style=boxed,basicstyle=\small]
  ...
  offload!(vec_add, &a, &b, &mut c, 0);
  offload!(vec_add, &a, &b, &mut c, 1);
  offload!(vec_add, &a, &b, &mut c, 2);
  offload!(vec_add, &a, &b, &mut c, 3);
  ...
\end{lstlisting}
Here we can clearly cancel out all intermediate data transfers that were generated in the lowering of our intrinsics.

\paragraph{Further experiments}
The benchmark that is the least amenable to our optimization experiments is the ``Energy'' benchmark in RAJAPerf~\cite{RajaPerf}.
It consists of six different kernels which share a large number of arguments. 
However, each kernel has a few unique arguments, and such an optimization would need a heuristic for whether it is reasonable to preload all of the data.
If applied, the transformation would increase the peak memory consumption on the GPU, potentially significantly, so we have therefore not prototyped it.

\paragraph{Rust-based performance improvements}
Most of our optimizations target the convenient offload interface and are therefore prototyped in LLVM, on which rustc already relies for many performance optimizations. 
For explicit data movement, however, rustc could also help through Clippy, the official Rust linter.
In particular, a future lint could warn when a preload call is placed unnecessarily late, leaving a lot of wasted time between the last CPU usage and the start of our Host to Device transfer.

%% file: sections/evaluation.tex
\input{sections/benchmark_data}

\section{Evaluation}\label{sec:benchmarks}
To show the portability and efficiency of our Rust GPU solution, we ported a subset of RAJAPerf~\cite{RajaPerf} to pure Rust and evaluated it against the original variants. RAJAPerf is the benchmark suite of RAJA~\cite{Raja}, a portable C++ framework for expressing loop-level parallelism across accelerators using different backends such as CUDA, HIP, and OpenMP.
Its kernels are derived from HPC applications and are commonly used to evaluate backend performance.
RAJA and its benchmarks are useful for our study as they allow us to compare our Rust Offload against the same RAJA-based code with different high-performance backends.

We ran the benchmarks on three different servers, using an AMD MI250X GPU, NVIDIA H100 GPU, and an NVIDIA RTX A2000 GPU.
Our extended Rust compiler is based on LLVM 23.1.0-rc1.

We evaluate five aspects: (1) kernel times, (2) memory transfer size, (3) total runtime, (4) fast-math impact, and (5) kernel register usage.

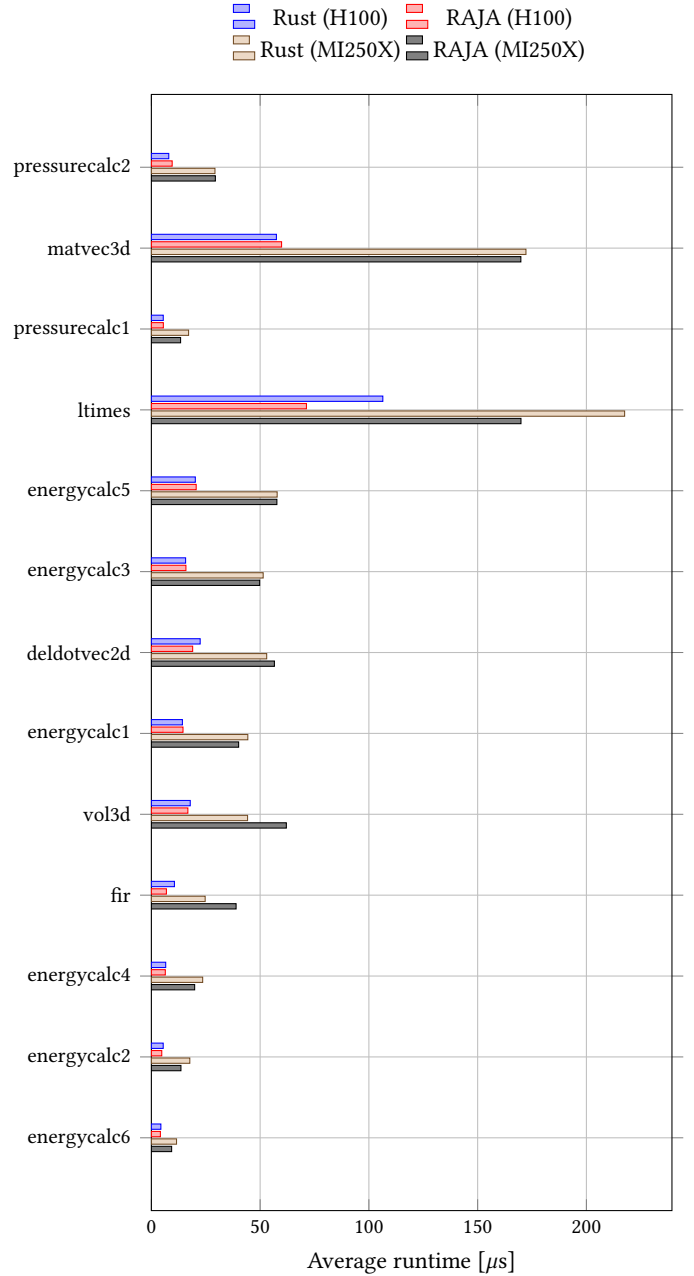
\begin{figure}[t]
\centering
\begin{tikzpicture}
\begin{axis}[
    xbar,
    width=\linewidth,
    height=0.72\textheight,
    bar width=2.0pt,
    xmin=0,
    enlarge y limits={abs=0.9},
    xlabel={Average runtime [$\mu$s]},
    scaled x ticks=false,
    ytick=data,
    yticklabels from table={\profilingKernelTimeMatrix}{Benchmark},
    y dir=reverse,
    tick label style={font=\footnotesize},
    label style={font=\small},
    legend columns=2,
    legend style={
        at={(0.5,1.02)},
        anchor=south,
        draw=none,
        fill=none,
        font=\small
    },
    xmajorgrids=true,
    ymajorgrids=false,
    grid=major,
]

\addplot+[
    xbar,
    bar shift=4.2pt,
] table[
    x expr={\thisrow{RustNs}/1000},
    y expr=\coordindex
] {\profilingKernelTimeMatrix};

\addplot+[
    xbar,
    bar shift=1.4pt,
] table[
    x expr={\thisrow{RajaNs}/1000},
    y expr=\coordindex
] {\profilingKernelTimeMatrix};

\addplot+[
    xbar,
    bar shift=-1.4pt,
] table[
    x expr={\thisrow{RustNs}/1000},
    y expr=\coordindex
] {\profilingTioga};

\addplot+[
    xbar,
    bar shift=-4.2pt,
] table[
    x expr={\thisrow{RajaNs}/1000},
    y expr=\coordindex
] {\profilingTioga};

\legend{
    Rust (H100),
    RAJA (H100),
    Rust (MI250X),
    RAJA (MI250X)
}

\end{axis}
\end{tikzpicture}

\caption{Average kernel runtime comparison between Rust and RAJA on H100 and MI250X.}
\label{fig:kernel-times}
\end{figure}

\paragraph{Kernel Times}
Figure~\ref{fig:kernel-times} shows the kernel times for Rust and RAJA variants on MI250X and H100. Rust kernels perform similarly to RAJA kernels except in the FIR and LTIMES benchmarks, where Rust shows slower kernel times.
Both are simple micro-benchmarks where each thread repeatedly performs a small number of multiply/add operations, with FIR using a compile-time loop count and LTIMES taking the iteration count as an argument.
Since these kernels are small and sensitive to unrolling decisions, larger differences are not surprising.

\paragraph{Memory Transfer Sizes}
While kernel times are similar, memory transfer behaviour differs between Rust and RAJA on both AMD and NVIDIA GPUs.
On H100 and across all benchmarks, RAJA performs slightly more and larger host-to-device transfers than Rust: 55 vs. 53 transfers, totalling 468 MB vs. 423 MB.
After the kernels, both perform 9 device-to-host transfers, but RAJA transfers 99 MB compared to Rust’s 69 MB.
Rust moves less data overall, but this does not translate into lower transfer time: its transfers take 46 ms compared to 16 ms for RAJA.
We suspect differences in memory kinds and asynchronous transfers to be the cause of the slowdown.

\begin{figure}[t]
\centering

\begin{tikzpicture}
\begin{axis}[
    ybar,
    ymode=log,
    log basis y=10,
    width=\columnwidth,
    height=55mm,
    bar width=2.6pt,
    ymin=4,
    ymax=2500,
    ylabel={Mean runtime (ms, log scale)},
    xmin=-0.5,
    xmax=6.5,
    xtick={0,1,2,3,4,5,6},
    xticklabels={
        DEL\_DOT\_VEC\_2D,
        ENERGY,
        FIR,
        LTIMES,
        MATVEC\_3D\_STENCIL,
        PRESSURE,
        VOL3D
    },
    xticklabel style={
        rotate=35,
        anchor=east,
        font=\scriptsize
    },
    ytick={5,10,20,50,100,200,500,1000,2000},
    tick label style={font=\scriptsize},
    label style={font=\small},
    legend style={
        at={(0.5,1.03)},
        anchor=south,
        legend columns=3,
        font=\scriptsize,
        /tikz/every even column/.append style={column sep=5pt}
    },
    enlarge x limits=0.08,
    unbounded coords=discard,
]

\addplot table[
    x expr=\coordindex,
    y expr=1000*\thisrow{BaseSeq}
] {\rajadatatioga};

\addplot table[
    x expr=\coordindex,
    y expr=1000*\thisrow{RustOffloadNaive}
] {\rustdatatioga};

\addplot table[
    x expr=\coordindex,
    y expr=1000*\thisrow{RustOffload}
] {\rustdatatioga};

\addplot table[
    x expr=\coordindex,
    y expr=1000*\thisrow{BaseHip}
] {\rajadatatioga};

\addplot table[
    x expr=\coordindex,
    y expr=1000*\thisrow{RajaHip}
] {\rajadatatioga};

\legend{
    Base\_Seq,
    Rust Interface A,
    Rust Interface C,
    Base\_HIP,
    RAJA\_HIP
}

\end{axis}
\end{tikzpicture}

\caption{
Mean benchmark runtime of Rust and RAJAPerf implementations on an AMD MI250X GPU.
}
\label{fig:runtime-tioga}
\end{figure}
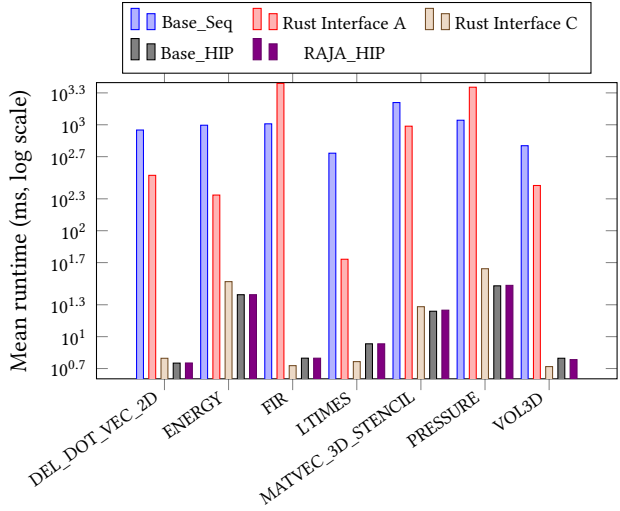

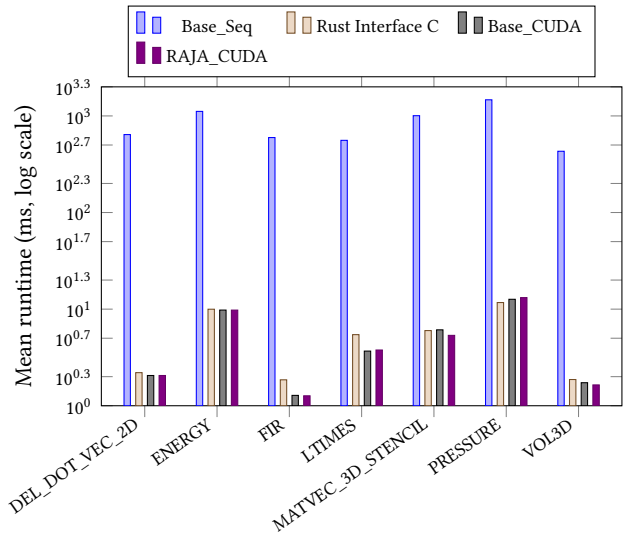
\begin{figure}[t]
\centering
\begin{tikzpicture}
\begin{axis}[
    ybar,
    ymode=log,
    log basis y=10,
    width=\columnwidth,
    height=58mm,
    bar width=2.4pt,
    ymin=1,
    ymax=2000,
    ylabel={Mean runtime (ms, log scale)},
    xmin=-0.6,
    xmax=6.6,
    xtick={0,1,2,3,4,5,6},
    xticklabels={
        DEL\_DOT\_VEC\_2D,
        ENERGY,
        FIR,
        LTIMES,
        MATVEC\_3D\_STENCIL,
        PRESSURE,
        VOL3D
    },
    xticklabel style={
        rotate=35,
        anchor=east,
        font=\scriptsize
    },
    ytick={1,2,5,10,20,50,100,200,500,1000,2000},
    tick label style={font=\scriptsize},
    label style={font=\small},
    legend style={
        at={(0.5,1.03)},
        anchor=south,
        legend columns=3,
        font=\scriptsize,
        /tikz/every even column/.append style={column sep=5pt}
    },
    enlarge x limits=false,
    unbounded coords=discard,
]

\addplot+ table[
    x expr=\coordindex,
    y expr={1000*\thisrow{BaseSeq}}
] {\rajadatamatrix};

\pgfplotsset{cycle list shift=1}

\addplot+ table[
    x expr=\coordindex,
    y expr={1000*\thisrow{RustOffload}}
] {\rustdatamatrix};


\addplot+ table[
    x expr=\coordindex,
    y expr={1000*\thisrow{BaseCUDA}}
] {\rajadatamatrix};

\addplot+ table[
    x expr=\coordindex,
    y expr={1000*\thisrow{RajaCUDA}}
] {\rajadatamatrix};

\legend{
    Base\_Seq,
    Rust Interface C,
    Base\_CUDA,
    RAJA\_CUDA
}
\end{axis}
\end{tikzpicture}
\caption{
Mean benchmark runtime of Rust and RAJAPerf implementations on an H100 NVIDIA GPU.
}
\label{fig:runtime-matrix}
\end{figure}

\paragraph{Benchmark Runtimes}
Figure~\ref{fig:runtime-tioga} and~\ref{fig:runtime-matrix} show the benchmark runtime for Rust and RAJA on MI250X and H100, respectively. These mainly include the kernel launch time, the kernel time, and the synchronization time.
As per RAJAPerf default, each benchmark kernel is launched in a loop, between 50 and 700 times, and the total runtime of this loop is then reported.
Memory transfer times are excluded by RAJAPerf, as transfers are generally done at the begin and end of the program.
As a smoke test, we add a naive Rust Offload implementation in Figure~\ref{fig:runtime-tioga} based on interface A, which transfers data once per kernel launch instead of once per benchmark. 
On our MI250X, this naive implementation can be over 400$\times$ slower than our optimized Rust implementation, which highlights the need for our optimizations described in \ref{sec:optimizations} to close this gap.

On MI250X (Figure~\ref{fig:runtime-tioga}), Rust is between 32\% faster and 43\% slower when measuring the whole runtime and not individual kernel timings.
On H100 (Figure~\ref{fig:runtime-matrix}), Rust is between 11\% faster and 46\% slower than the base CUDA implementation.
The biggest differences in favour of BaseCuda can be seen on the FIR and LTIMES benchmark, where Rust is 44\% and 46\% slower, respectively. These two benchmarks also give Rust the biggest advantage over BaseHIP, with 15\% and 32\% respectively. Both benchmarks consist only of a very small loop with few instructions, and as such different unroll decisions between the three compilers can have a major impact.

\begin{figure}[t]
\centering
\begin{tikzpicture}
\begin{axis}[
    xbar,
    width=\linewidth,
    height=0.72\textheight,
    bar width=2.0pt,
    xmin=0,
    enlarge y limits={abs=0.9},
    xlabel={Average runtime [$\mu$s]},
    scaled x ticks=false,
    ytick=data,
    yticklabels from table={\profilingKernelTimeUofT}{Benchmark},
    y dir=reverse,
    tick label style={font=\footnotesize},
    label style={font=\small},
    legend columns=2,
    legend style={
        at={(0.5,1.02)},
        anchor=south,
        draw=none,
        fill=none,
        font=\small
    },
    xmajorgrids=true,
    ymajorgrids=false,
    grid=major,
]

\addplot+[
    xbar,
    bar shift=4.2pt,
] table[
    x expr={\thisrow{RustNs}/1000},
    y expr=\coordindex
] {\profilingKernelTimeUofT};

\addplot+[
    xbar,
    bar shift=1.4pt,
] table[
    x expr={\thisrow{RustAlgebraicNs}/1000},
    y expr=\coordindex
] {\profilingKernelTimeUofT};

\addplot+[
    xbar,
    bar shift=-1.4pt,
] table[
    x expr={\thisrow{BaseCUDANs}/1000},
    y expr=\coordindex
] {\profilingKernelTimeUofT};

\addplot+[
    xbar,
    bar shift=-4.2pt,
] table[
    x expr={\thisrow{RAJACUDANs}/1000},
    y expr=\coordindex
] {\profilingKernelTimeUofT};

\legend{
    Rust,
    Rust (Algebraic Floats),
    Base\_CUDA,
    RAJA\_CUDA,
}

\end{axis}
\end{tikzpicture}

\caption{Average kernel runtime comparison between Rust and RAJA on NVIDIA RTX A2000.}
\label{fig:algebraic}
\end{figure}
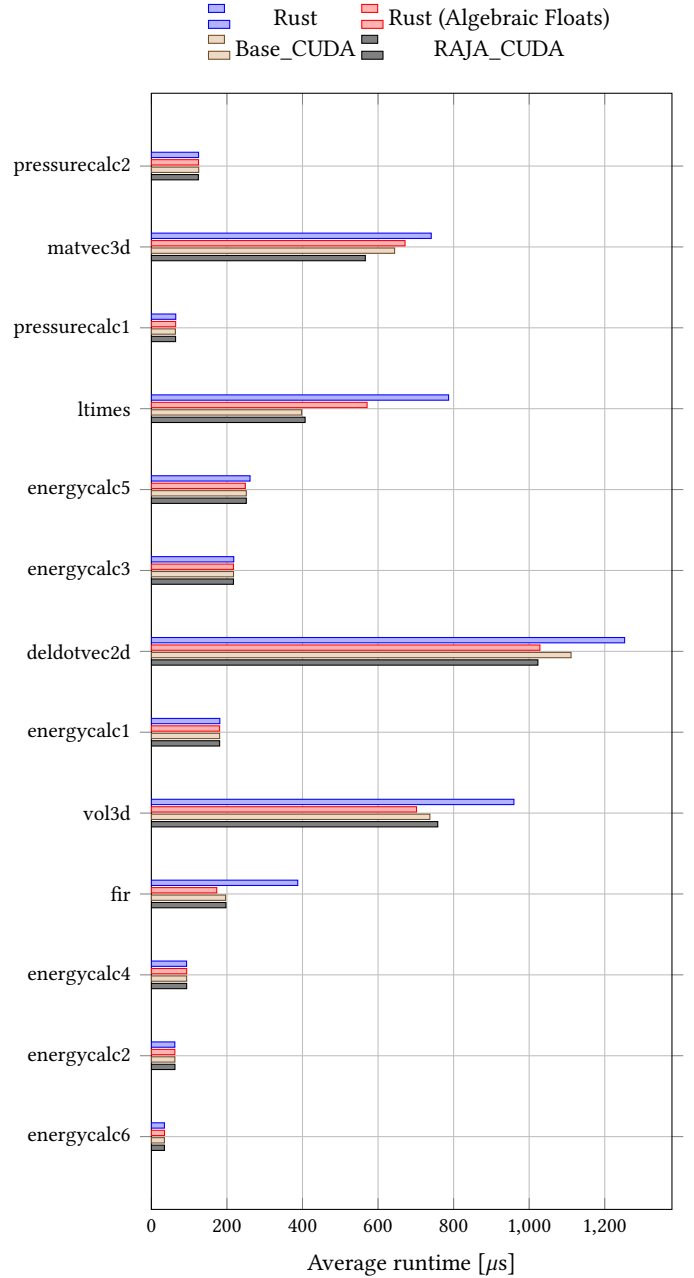

\paragraph{Fast-Math Impact}
To assess whether Rust kernel times benefit from common GPU optimization settings, we evaluate relaxed floating-point semantics. 
C++ GPU codes commonly use fast-math, which combines seven flags that enable optimizations based on weaker floating-point guarantees.
Rust does not allow users to enable fast-math directly, since they include ``no NaN'' (nnan) and ``no infinity'' (ninf) assumptions that could trigger UB in safe Rust code. 
Instead, Rust provides experimental algebraic~\cite{AlgebraicFloats} floating-point operations, which expose most of the relevant optimization opportunities while excluding the nnan and ninf assumptions.
As Figure~\ref{fig:algebraic} shows, on an RTX A2000, algebraic floats provide a 2$\times$ speedup on FIR and improve \lstinline{DEL_DOT_VEC_2D}, \lstinline{VOL3D}, and \lstinline{MATVEC3D} by about 20\%.
The FIR kernel is a trivial loop and already unrolled in the case of normal floats. Algebraic floats allow LLVM to further vectorize it with a vector-width of 4, explaining the large impact. Algebraic floats do not have a relevant impact on other kernels. 
Using algebraic floats on the MI250X has not shown significant performance improvements.

\paragraph{Register Usage}
We primarily care about the pure kernel times, but register usage is another indicator to confirm that we generate efficient LLVM-IR.
On RTX 2070, the average register usage of Rust is 33, while the RAJA-CUDA solution averages 28 across the 13 implemented RAJAPerf kernels.
Using fast-math via algebraic floats in Rust only results in a minimal increase in two of the kernels.
The overall slightly higher register usage could be caused by additional bounds-checking in the Rust code.
In Rust CPU code, bounds checking is usually avoided by iterating over all elements in an array instead of indexing into it.
GPU code uses explicit indexing based on the thread and block indices instead, so Rust must verify that the access is within bounds.
In RAJAPerf, most array, block, and thread dimensions are known at compile time, and we have not measured any runtime impact of bounds-checking on Rust GPU kernels. 

%% file: sections/benchmark_data.tex
\pgfplotstableread[row sep=newline, col sep=space]{
App RustOffloadNaive RustOffload
DEL_DOT_VEC_2D           0.333227  0.006250 
ENERGY                   0.217053  0.033069 
FIR                      2.456091  0.005327 
LTIMES                   0.053783  0.005814 
MATVEC_3D_STENCIL        0.969724  0.019183 
PRESSURE                 2.261837  0.043775 
VOL3D                    0.267186  0.005218 
}\rustdatatioga

\pgfplotstableread[row sep=newline, col sep=space]{
BaseSeq BaseHip RajaHip
0.892041    0.005627   0.005642  
0.990183    0.024888   0.024894  
1.020672    0.006265   0.006262  
0.539111    0.008560   0.008556  
1.621313    0.017373   0.017783  
1.103730    0.030184   0.030494  
0.634521    0.006255   0.006071  
}\rajadatatioga

\pgfplotstableread[row sep=newline, col sep=space]{
App RustOffload
DEL_DOT_VEC_2D           0.002201 
ENERGY                   0.009981 
FIR                      0.001855 
LTIMES                   0.005446 
MATVEC_3D_STENCIL        0.005999 
PRESSURE                 0.011703 
VOL3D                    0.001868 
}\rustdatamatrix

\pgfplotstableread[row sep=newline, col sep=comma]{
BaseSeq,BaseCUDA,RajaCUDA 
0.641180  ,  0.002056 ,  0.002056 
1.114070  ,  0.009780 ,  0.009772 
0.597219  ,  0.001280 ,  0.001267 
0.558844  ,  0.003679 ,  0.003775
1.005831  ,  0.006105 ,  0.005352 
1.468593  ,  0.012647 ,  0.013159 
0.430756  ,  0.001732 ,  0.001643 
}\rajadatamatrix

\pgfplotstableread[row sep=newline, col sep=comma]{
App,RustOffload,BaseCUDA,SlowdownRustOverBaseCUDA
DEL_DOT_VEC_2D,0.006517,0.002056,3.169747081712062
ENERGY,0.041664,0.00978,4.2601226993865025
FIR,0.007978,0.00128,6.2328125
LTIMES,0.007281,0.003679,1.9790703995650993
MATVEC_3D_STENCIL,0.01101,0.006105,1.8034398034398036
PRESSURE,0.062659,0.012647,4.95445560211908
VOL3D,0.005682,0.001732,3.2806004618937648
}\H

\pgfplotstableread[row sep=newline, col sep=space]{
Total-Time(ns)  Instances  Avg(ns)  Med(ns)   Min(ns)  Max(ns)  StdDev(ns)       Name
5,748,687        100   57,486.9   57,552.5    56,097    58,848        587.1  matvec3d
5,596,017        700    7,994.3    7,968.0     7,616     8,800        170.6  pressurecalc2
5,320,429         50  106,408.6  106,400.0   106,336   106,528         43.2  ltimes
3,825,930        700    5,465.6    5,440.0     5,248     6,080        133.0  pressurecalc1
2,626,694        130   20,205.3   20,080.5    19,840    21,088        332.2  energycalc5
2,243,782        100   22,437.8   22,400.0    19,968    23,520        373.8  deldotvec2d
2,045,702        130   15,736.2   15,648.0    15,392    16,768        288.4  energycalc3
1,854,275        130   14,263.7   14,144.0    13,952    15,200        285.4  energycalc1
1,787,013        100   17,870.1   17,856.0    15,520    18,560        324.9  vol3d
1,696,227        160   10,601.4   10,592.0    10,528    11,072         53.4  fir
  856,323        130    6,587.1    6,592.0     6,400     7,008         89.6  energycalc4
  707,298        130    5,440.8    5,440.0     5,344     5,760         59.8  energycalc2
  567,585        130    4,366.0    4,352.0     4,256     4,736         94.1  energycalc6
}\profilingRustMatrix

\pgfplotstableread[row sep=newline, col sep=space]{
Total-Time(ns)  Instances  Avg(ns)  Med(ns)  Min(ns)  Max(ns)  StdDev(ns)  Name
6,694,546        700   9,563.6   9,568.0     9,088    10,528        190.5  pressurecalc2
5,986,032        100  59,860.3  59,872.0    57,952    61,217        619.8  matvec_3d
3,852,426        700   5,503.5   5,472.0     5,248     6,400        144.6  pressurecalc1
3,564,330         50  71,286.6  71,264.5    69,664    72,256        506.1  ltimes
2,678,214        130  20,601.6  20,352.0    20,064    21,760        517.4  energycalc5
2,064,709        130  15,882.4  15,648.0    15,232    17,184        513.1  energycalc3
1,898,310        100  18,983.1  18,848.0    17,184    20,385        433.0  deldotvec2d
1,890,630        130  14,543.3  14,304.0    14,144    15,585        473.0  energycalc1
1,677,893        100  16,778.9  16,736.0    15,616    17,664        285.6  vol3d
1,106,466        160   6,915.4   6,912.0     6,816     7,616         74.8  fir
  832,802        130   6,406.2   6,400.0     6,176     7,008        108.5  energycalc4
  621,121        130   4,777.9   4,768.0     4,576     5,568        126.7  energycalc2
  538,688        130   4,143.8   4,128.0     4,064     4,352         50.1  energycalc6
}\profilingRajaMatrix

\pgfplotstableread[row sep=newline, col sep=comma]{
"Name","Calls","TotalDurationNs","AverageNs","Percentage"
"pressurecalc2",700,20291284,28987,20.191708589092887
"matvec3dstencil",100,17821319,178213,17.733864447477266
"pressurecalc1",700,11656296,16651,11.599095062698304
"ltimes",50,10869241,217384,10.815902377425726
"energycalc5",130,7407441,56980,7.371090467360213
"energycalc3",130,6479274,49840,6.447478260956095
"energycalc1",130,5561344,42779,5.534052818525442
"del_dot_vec_2d",100,5316861,53168,5.290769569866205
"vol3d",100,4337968,43379,4.316680291144222
"fir",160,3924209,24526,3.9049517305408377
"energycalc4",130,3024511,23265,3.0096688182229334
"energycalc2",130,2301786,17706,2.290490446363757
"energycalc6",130,1501616,11550,1.4942471203261118
}\profilingRustTioga

\pgfplotstableread[row sep=newline, col sep=comma]{
"Name","Calls","TotalDurationNs","AverageNs","Percentage"
pressurecalc2,700,20593622,29419,20.937010750804372
matvec_3d,100,16986867,169868,17.270114844367058
pressurecalc1,700,9402960,13432,9.559749839507761
ltimes,50,8495914,169918,8.637579283329051
energycalc5,130,7501185,57701,7.626263655260473
energycalc3,130,6471413,49780,6.579320702006435
fir,160,6231895,38949,6.335808854454259
vol3d,100,6205173,62051,6.308641277945231
deldotvec2d,100,5656209,56562,5.750523567044032
energycalc1,130,5215086,40116,5.3020450529959895
energycalc4,130,2586743,19898,2.629875696493213
energycalc2,130,1766580,13589,1.7960368725888038
energycalc6,130,1211049,9315,1.2312426601183069
}\profilingRajaTioga

\pgfplotstableread[row sep=newline, col sep=comma]{
Benchmark,RajaNs,RustNs,Speedup
pressurecalc2,9563.6,7994.3,1.1963023654353728
matvec3d,59860.3,57486.9,1.0412859277504962
pressurecalc1,5503.5,5465.6,1.0069342798594847
ltimes,71286.6,106408.6,0.6699326934101191
energycalc5,20601.6,20205.3,1.0196136657213701
energycalc3,15882.4,15736.2,1.00929068008795
deldotvec2d,18983.1,22437.8,0.8460321421886281
energycalc1,14543.3,14263.7,1.0196022070009885
vol3d,16778.9,17870.1,0.9389371072349905
fir,6915.4,10601.4,0.6523100722546079
energycalc4,6406.2,6587.1,0.9725372318622761
energycalc2,4777.9,5440.8,0.8781612998088515
energycalc6,4143.8,4366.0,0.9491067338524966
}\profilingKernelTimeMatrix

\pgfplotstableread[row sep=newline, col sep=comma]{
Benchmark,RajaNs,RustNs,Speedup
pressurecalc2,29419.0,29236.0,1.00625940621152
matvec3d,169868.0,172212.0,0.9863888695329013
pressurecalc1,13432.0,17108.0,0.7851297638531681
ltimes,169918.0,217572.0,0.7809736546982149
energycalc5,57701.0,57832.0,0.9977348180937889
energycalc3,49780.0,51404.0,0.9684071278499727
deldotvec2d,56562.0,53017.0,1.066865345077994
energycalc1,40116.0,44339.0,0.9047565348790004
vol3d,62051.0,44224.0,1.4031069102749638
fir,38949.0,24730.0,1.5749696724625961
energycalc4,19898.0,23584.0,0.8437075983717774
energycalc2,13589.0,17668.0,0.7691306316504415
energycalc6,9315.0,11564.0,0.8055171221030785
}\profilingTioga

\pgfplotstableread[row sep=newline, col sep=comma]{
Benchmark,RustNs,RustAlgebraicNs,BaseCUDANs,RAJACUDANs,Speedup
pressurecalc2,124869.0,124772.1,125322.8,124372.0,
matvec3d,740781.9,671459.0,643838.0,566294.3,
pressurecalc1,64543.6,64283.5,63612.3,64082.2,
ltimes,786712.3,570984.6,397597.8,406900.8,
energycalc5,261171.1,248742.2,251317.5,251285.5,
energycalc3,218042.0,217393.4,217404.1,217338.9,
deldotvec2d,1252695.8,1028520.5,1110918.8,1023101.3,
energycalc1,181135.8,180655.3,180708.5,180684.4,
vol3d,959662.5,701820.0,737184.2,758056.1,
fir,387479.9,173078.7,196399.9,197350.6,
energycalc4,93215.3,93450.6,93465.3,93470.9,
energycalc2,62116.0,62170.9,62249.6,62424.1,
energycalc6,34537.7,34789.3,34509.5,34568.4,
}\profilingKernelTimeUofT